\documentclass{aa}  
\usepackage{graphicx}
\usepackage{txfonts}
\begin{document}
   \title{El Roque de Los Muchachos Site Characteristics. II.}

   \subtitle{Analysis of Wind, Relative Humidity and Air Pressure}

   \author{G. Lombardi\inst{1,2}
          \and
          V. Zitelli\inst{2}
          \and
          S. Ortolani\inst{3}
          \and
          M. Pedani\inst{4}
          }

   \offprints{G. Lombardi}

   \institute{University of Bologna, Department of Astronomy, via Ranzani 1, I-40127 Bologna, Italy\\
    \email{gianluca.lombardi@oabo.inaf.it}
         \and
             INAF - Bologna Astronomical Observatory, via Ranzani 1, I-40127 Bologna, Italy\\
             \email{valentina.zitelli@oabo.inaf.it}
         \and
             University of Padova, Department of Astronomy, vicolo dell'Osservatorio 2, I-35122, Padova, Italy\\
             \email{sergio.ortolani@unipd.it}
         \and
             Fundaci\'on Galileo Galilei and Telescopio Nazionale Galileo, PO Box 565, E-38700, S/C de La Palma, Tenerife, Spain\\
             \email{pedani@tng.iac.es}
             }

   \date{Accepted for publication in PASP}

 
  \abstract
   {In this paper we present an analysis of wind speed, wind direction, relative humidity and air pressure taken at Telescopio Nazionale Galileo (TNG), Carlsberg Automatic Meridian Circle (CAMC; now called Carlsberg Meridian Telescope) and Nordic Optical Telescope (NOT) at Observatorio del Roque de Los Muchachos, in the Canary Islands.}
   {Data are also compared in order to check local variations and both long term and short term trends of the microclimate. Furthermore, influence of wind speed on the astronomical seeing is estimated to the aim to better understand the influence of wide scale parameters on local meteorological data.}
   {The analysis is done using a statistical approach. From each long series of data we compute the hourly, daily and monthly averages. A particular care is used to minimize any effect due to biases in case of missing of data. Wind direction is estimated by the annual percentage of the time in which the wind come from fixed directions.}
   {We found that relative humidity presents a negative correlation with temperature and pressure, while pressure is correlated to the temperature. The three telescopes show different prevailing wind direction, wind speed, relative humidity and air pressure confirming differences in local microclimate. We found that seeing deteriorates when wind speed is lower than 3.3 m s$^{-1}$. Comparison in terms of wind speed and high relative humidity ($> 90\%$) shows that TNG seems to have optimal observational conditions with respect to CAMC and NOT. Air pressure analysis shows that ORM is dominated by high pressure, that means prevailing stable good weather, as is to be expected from anticyclonic conditions. Finally, short time variations of pressure anticipate temperature variations tipically by 2-3 hours, this property vanishes in time scales higher than some hours and disappear in longer time scales.}
   {}


   \maketitle
%

\section{Introduction}
Since 1970, the Observatorio del Roque de Los Muchachos (ORM) located at La Palma Island (Canaries) hosts Europe's main astronomical telescopes. The very good astronomical conditions of the island are mainly due to a stable subsiding maritime air mass, which typically prompts the telescopes to be placed near the top of the mountain, well above the inversion layer occurring in the range between 800 and 1200 m (McInnes \& Walker \cite{mcinnes}).\\
All the telescopes are located along the northern edge of the Caldera de Taburiente, at the northwest side of La Palma Island, where the irregular shapes produce a complex orography, and the crowdedness of the top, due to the presence of all the astronomical observatories, suggests a possible modification of the local microclimate, making it difficult to foresee in advance the precise local meteorological parameters.\\
In Lombardi et al. \cite{lombardi} (hereafter Paper I) we have presented for the first time an analysis of temperature obtained from local meteorological towers, plus environmental conditions measured at two telescopes at ORM. Meteorological data from the
Telescopio Nazionale Galileo (TNG) and the Carlsberg Meridian Telescope (formerly the Carlsberg Automatic Meridian Circle,
CAMC) have been compared in order to check local variations in meteorological conditions due to the tamperature variations. We also investigated the influence of temperature on astronomical seeing at TNG concluding that seeing deteriorates when the temperature around the dome at the same height of the primary mirror of the telescope is at least 0.6 deg higher than the temperature below that height.\\
In this paper we continue the analysis of La Palma climate, presenting for the first time an accurate study of wind speed and direction, relative humidity and air pressure. In order to give a more detailed comparison between different sites at ORM, data from Nordic Optical Telescope (NOT) have also been added to those of TNG and CAMC. An analysis in terms of image quality at the telescopes will be discussed. The structure of the paper is the following: in \S 2 the databases analysis is discussed in order to explain the used statistics; in \S 3 we analyze the influence of wind speed on the astronomical seeing, moreover the behaviour of the wind both in speed and direction at the three telescope sites is evaluated; in \S 4 the relative humidity is discussed and its comparison with the downtime at TNG and CAMC is shown; in \S 5 air pressure and its influence on weather conditions at ORM is discussed.\\
In the following sections we will often make use of temperatures ($T$) obtained in Paper I in order to compare the meteorological parameters.
\section{Data analysis}\label{data-analysis}
Meteorological data are sampled from TNG, CAMC and NOT meteo stations which have different locations in the ORM. It is important to note that all the data series provided by each meteo sensor have to be considered in local time. The three sites are located well above the inversion layer as shown in Table \ref{telescopes} that lists positions and altitudes above the sea level of the telescopes.\\
   \begin{table}[t]
     \begin{center}\scriptsize
       \caption[]{Locations and altitudes of TNG, CAMC and NOT.}
    \label{telescopes}
        \begin{tabular}{l | c | c | l}
           \hline
            \noalign{\smallskip}
            & Latitude & Longitude & Altitude a.s.l. [m]\\
            \noalign{\smallskip}
            \hline
            \noalign{\smallskip}
TNG & 28$^{\circ}$ 45' 28.3'' N & 17$^{\circ}$ 53' 37.9'' W & 2387 (Elevation axis)\\
CAMC & 28$^{\circ}$ 45' 36.0'' N & 17$^{\circ}$ 52' 57.0'' W & 2326 (Dome floor)\\
NOT & 28$^{\circ}$ 45' 26.2'' N & 17$^{\circ}$ 53' 06.3'' W & 2382 (Dome floor)\\
            \noalign{\smallskip}
            \hline
         \end{tabular}
         \end{center}
   \end{table}
The TNG meteorological tower is a 15 m tall iron structure located about 100 m away from TNG building. The data are regularly sent from the tower to the TNG annex building using an optical fiber link since 1998 March 27 (Porceddu et al. \cite{porceddu}; Paper I).\\
The CAMC carried out regular meteorological observations in the period of 1984 May 13 to 2005 March 31, and the records are more or less continuous in that period. For the years 1984, 1985, and 1986, meteorological readings are only available each 30 minute of intervals. Starting in 1987 January, readings were made each 5 minute of intervals during day and night. In the beginning of December 1994, all readings were made each 20 s of intervals and were then averaged using a time scale of 5 minutes\footnote{http://www.ast.cam.ac.uk}.\\
The NOT provides a complete archive since 2 March 1997. The data are available in regular readings done every 5 minutes\footnote{http://www.not.iac.es}.\\
Following the same procedures as described in Paper I we have analyzed wind speed, relative humidity and air pressure. From each raw data series, we compute the hourly averages, and then from each set of these, we compute the monthly averages. Particular care was taken to minimize any effects due to biases in cases of missing of data, which typically occurred in wintertime. In case of missing values, we take into account the averages obtained from two corresponding months in other years in which the values of the adjacent months are similar. A more detailed description of the adopted procedure can be found in Paper I.\\
Custom changes to the general analysis method for each meteorological parameter have been introduced. The following items describe these changes plus environamental informations about the parameters.
\begin{description}
	\item{$1.$} \textbf{Wind vector ($\vec{V}$)}\\
	Wind vector can be assumed as $\vec{V} = w_{sp}\vec{w}_{dir}$, where $w_{sp}$ and $\vec{w}_{dir}$ are respectively wind speed and wind direction. The wind speed is measured in m s$^{-1}$, while wind direction in degrees (North is represented with 0$^{\circ}$, East with 90$^{\circ}$). TNG sensor is placed at the top of the meteo tower and has an accuracy better than $2\%$ for $w_{sp}$ and $\pm 3^{\circ}$ for $\vec{w}_{dir}$. CAMC sensor was placed 6 m above the ground until 1991 May 16, then it was moved 10.5 m above the ground. It has a wind speed accuracy of $\pm 1\%$ below 20 m s$^{-1}$ and $\pm 2\%$ above 20 m s$^{-1}$, while wind direction is provided with a $\pm 5^{\circ}$ accuracy. NOT sensor has an accuracy better than $2\%$ for $w_{sp}$ and better than $5^{\circ}$ for $\vec{w}_{dir}$.\\
	Wind vector has been analyzed considering its daytime and nighttime behaviour. Following Paper I daytime data have been defined in the range 10:00$-$16:00 (local time), while nighttime data in the range 22:00$-$4:00 (local time). From each raw data series of $w_{sp}$ and $\vec{w}_{dir}$ we computed the hourly averages.\\
From each set of wind speed hourly averages, we computed the monthly averages and then the annual averages for both daytime and nighttime.\\
Daytime and nighttime wind direction statistics have been evaluated by calculating the annual percentage of hours in which the wind come from each direction $\vec{D}$. The wind rose has been divided into 8 mean directions (N, NE, E, SE, S, SW, W, NW) and the percentages of hours are calculated into intervals defined as [$\vec{D} - 22.5^{\circ}, \vec{D} + 22.5^{\circ}$[.
	\item{$2.$} \textbf{Relative Humidity ($RH$)}\\ 
	The relative humidity is the percentage of water vapour in the air with respect to the theoretical amount necessary to reach condensation at the same temperature.\\
	For the three telescopes the relative humidity sensors give an accuracy better than $2\%$. TNG sensor is placed 2 m above the ground on the meteo tower, while CAMC sensor was placed inside the dome until 1987 October 17 when it was  moved to the outside north-facing wall of the dome; finally, NOT sensor is placed outside the dome.\\
	Hourly, monthly and annual averages have been calculated in the usual manner. Also in this case, daytime (10:00$-$16:00) and nighttime (22:00$-$4:00) statistics have been considered. Entire day (00:00$-$24:00; local time) statistics have been calculated too.
	\item{$3.$} \textbf{Air Pressure ($P$)}\\
	Air pressure data are sampled with an accuray of $\pm 0.1$ hPa. The TNG sensor is placed 1 m above the ground on the meteo tower, the CAMC one is placed inside the dome, 1 m above the floor and NOT sensor is also placed inside the dome, 2 m above the floor.\\
\end{description}
\section{Wind}\label{wind}
\subsection{Wind and astronomical seeing}\label{wind-seeing}
\begin{figure}[t]
   \centering
   \includegraphics[width=8cm]{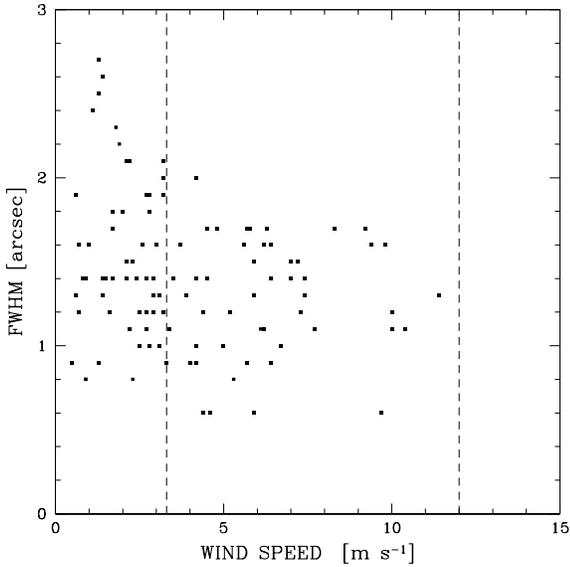}
   \caption{Seeing in $V$-band VS $w_{sp}$ at TNG. The two extremes for optimal observations are indicated.}
              \label{WSP-FWHM}
    \end{figure}

Wind speed is an important parameter because is linked to the optical turbulence ($C_{N}^{2}$) and to the wave-front coherence time (Geissler \& Masciadri \cite{geissler}). It is well known (Sarazin \cite{sarazin92}) that the effects of wind velocity are negligible for $w_{sp} \in$ [$w_{MIN}, w_{MAX}$[, where the two extremes are site dependent and $w_{MIN} > 0$.\\
In Paper I we found that the seeing, measured as FWHM of stellar image profiles obtained at the telescope, deteriorates when the temperature $T_{M}$ around the dome at the same height of the primary mirror of the telescope ($h_{M}$) is at least 0.6 deg higher than the temperature below that height (see Figure 8 in Paper I).\\
We make use of 118 images obtained with the image camera OIG (Optical Imager of Galileo) at TNG, pointed near the zenith (and corrected to the true zenith by a small amount), from 31 January to 4 February 2000. We computed FWHM of several stellar images in the $V$-band frames. The images have been processed following the standard procedure (bias subtraction and flat fielding) using IRAF packages. The images quality in term of FWHM is compared to the wind speed measured at the same starting UTs of each image.\\
Figure \ref{WSP-FWHM} displays the comparison between $w_{sp}$ and the FWHM. We see that the 50\% of the points is distribuited below $w_{sp} < 3.3$ m s$^{-1}$ (first vertical dashed line) where the median values of FHWM is 1.5 arcsec. Furthermore, for $w_{sp} \geq 3.3$ m s$^{-1}$ the distribution of the points shows a median value of 1.3 arcsec. This indicates that for $w_{sp} < 3.3$ m s$^{-1}$ the seeing deteriorates, so we can define $w_{MIN} = 3.3$ m s$^{-1}$. No observations are available for $w_{sp} > 12$ m s$^{-1}$ (second vertical dashed line). Sarazin \cite{sarazin92} shows in La Silla a limiting value of $w_{MAX} = 12$ m s$^{-1}$. Finally, we can conclude that TNG has optimal seeing conditions if $w_{sp} \in$ [3.3, 12[.\\
\begin{figure}[t]
   \centering
   \includegraphics[width=5.3cm]{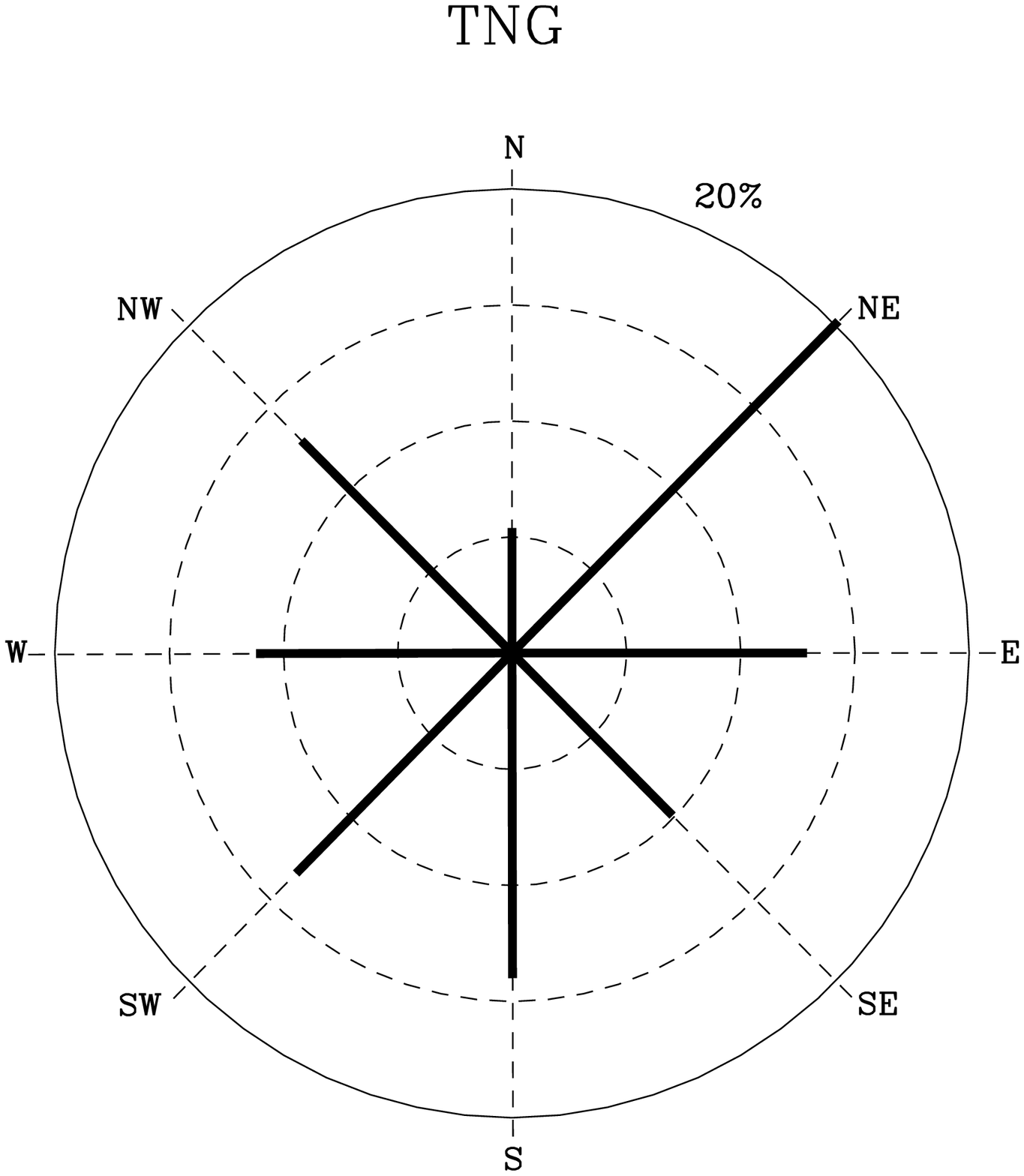}
   \includegraphics[width=5.3cm]{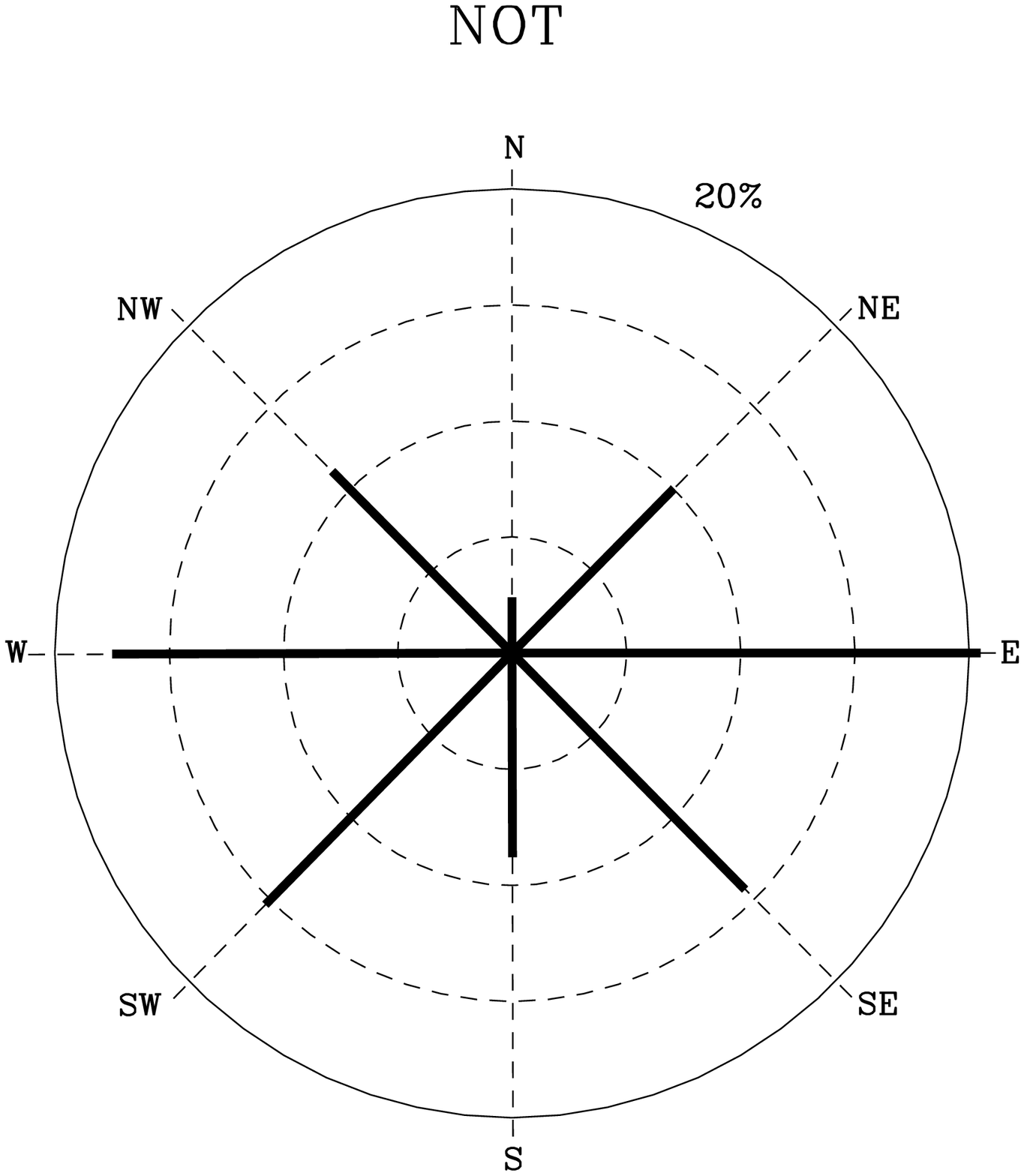}
   \includegraphics[width=5.3cm]{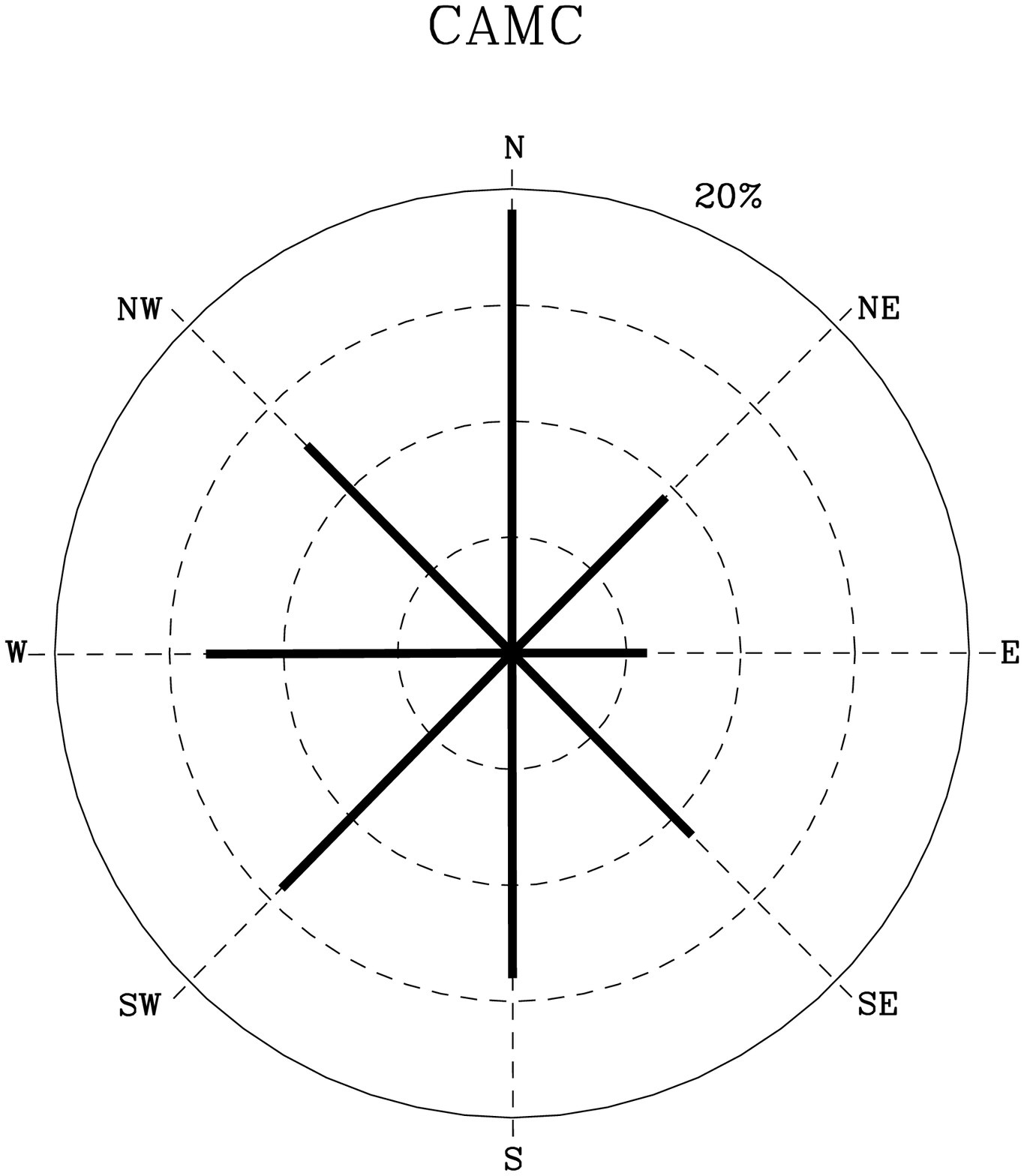}
   \caption{Nighttime wind roses for TNG, NOT and CAMC in the common period 1998-2004.}
              \label{wind-rose}
    \end{figure}

\subsection{Wind speed and direction}\label{wind-sub}
Trade winds are the main responsibles of the climate of Canary Islands. Font-Tullot \cite{font-tullot} affirms that the overall climate of the Canary Islands is determined by the trade winds which are 90\% prevalent in summer and 50\% prevalent in winter, with main direction from NW at the level of the observatories. Mahoney et al. \cite{mahoney} point out to a seasonal variation due to the Azores anticyclone which, together with the Canary Current, drives the trades in N-S direction.\\
Several authors analyzed the wind pattern and speed at ORM, but results are quite different. In the period 1971-1976 Brandt \& Righini \cite{brandt85} obtained a mean velocity of 6 m s$^{-1}$ and dominant wind direction NW with secondary peaks from NE to SE at ORM JOSO (Joint Organization for Solar Observations) sites. Daytime statistics by Brandt \& Woehl \cite{brandt82} for period 1978-1979 on the neighbouring slope of ORM shows distribution in wind direction with a prevailing component from E.\\
Nighttime data found by Mahoney et al. \cite{mahoney} at Gran Telescopio Canarias (GTC) site denote NE dominant wind direction. Further analysis by Jabiri et al. \cite{jabiri} at CAMC site in the period 1987-1995 shows a prevailing wind flow from N-NW during the day that change to N-NE in nighttime. Jabiri's wind speed analysis give a mean $w_{sp}$ of 2.8 m s$^{-1}$.\\
We can conclude that the wind direction significantly changes across the site.\\
In our analysis we make use of TNG and NOT 7 years long databases (1998-2004) and CAMC 20 years long database (1985-2004). The TNG sensor is placed on the top of the meteo tower at an altitude of 2370 m above sea level, while NOT and CAMC sensors are approximatively placed at the same height of the respective dome floors, that means 2380 m for the NOT sensor and 2325 m for the CAMC sensor.\\
Figure \ref{wind-rose} shows the nighttime wind roses for TNG, NOT and CAMC calculated taking into account the common period 1998-2004. More detailed informations about the behaviour of the wind direction in each yaer are given in Tables \ref{tng-night-wdir}, \ref{not-night-wdir} and \ref{camc-night-wdir}.\\
TNG shows a NE dominant mode in nighttime (see Table \ref{tng-night-wdir}) and a less evident prevailing wind direction from S to W in daytime. The mean wind speed is about 4.6 m s$^{-1}$, lower than the 6 m s$^{-1}$ found by Brandt \& Righini \cite{brandt85} and higher than the 2.8 m s$^{-1}$ of Jabiri et al. \cite{jabiri}. The maximum wind speed measured at TNG is of 26.9 m s$^{-1}$ in January 1999.\\
Dominant direction both in nighttime (Table \ref{not-night-wdir}) and daytime at NOT has two prevailing directions from W or E. NOT shows a mean $w_{sp}$ of 7.2 m s$^{-1}$, the highest measured at ORM. The maximum wind speed measured at NOT is 29.8 m s$^{-1}$ in December 2004.\\
CAMC shows a lower mean wind speed (2.2 m s$^{-1}$). This result is also lower than the 2.8 m s$^{-1}$ at CAMC found of Jabiri et al. \cite{jabiri}. The difference can be explained in term of statistics because our analysis uses a longer database. The maximum wind speed measured at CAMC is 18.3 m s$^{-1}$ in April 1987.\\
Wind direction for CAMC is very peculiar. Table \ref{camc-night-wdir} shows nighttime $\vec{w}_{dir}$ percentages per sector in nighttime over the two analyzed decades. Northern winds seems to oscillate with a period of 10 years, while winds from NW shows a similar oscillation in opposite phase. Anyway, there are no evidences of a prevailing direction, furthermore N and NW percentages become periodically comparable with those of other directions. The situation changes dramatically in 2004 when the percentage of wind from N increases steeply up to $71.8\%$. The recent behaviour of the wind at CAMC point out that a deeper analysis of such phenomenon will be needed in the future.\\
\begin{table}[t!]
     \begin{center}\scriptsize
       \caption[]{1998-2004 nighttime wind direction percentages for TNG.}
    \label{tng-night-wdir}
         \begin{tabular}{r | c | c | c | c | c | c | c | c}
	 \hline\hline
          \noalign{\smallskip}
           & N & NE & E & SE & S & SW & W & NW\\
          \noalign{\smallskip}
\hline
1998   &      8.4    &    22.6   &     11.2   &     11.8    &    14.6    &    10.4    &    10.3   &     10.7\\
1999   &      6.0    &    20.3   &     11.3   &      8.0    &    13.8    &    12.9    &    10.3   &     17.4\\
2000   &      4.1    &    19.7   &     11.4   &     11.3    &    14.1    &    13.3    &    10.9   &     15.2\\
2001   &      6.8    &    21.5   &     14.0   &      7.0    &    12.3    &    12.5    &    12.6   &     13.3\\
2002   &      4.2    &    18.9   &     14.3   &     10.0    &    14.8    &    14.0    &    13.2   &     10.6\\
2003   &      5.0    &    17.8   &     11.3   &      9.6    &    14.7    &    15.5    &    12.8   &     13.3\\
2004   &      4.6    &    21.2   &     15.6   &     11.1    &    13.3    &    14.2    &     8.3   &     11.7\\
\hline
\hline
         \end{tabular}
         \end{center}
\end{table}
\begin{table}[t!]
     \begin{center}\scriptsize
       \caption[]{1998-2004 nighttime wind direction percentages for NOT.}
    \label{not-night-wdir}
         \begin{tabular}{r | c | c | c | c | c | c | c | c}
	 \hline\hline
          \noalign{\smallskip}
           & N & NE & E & SE & S & SW & W & NW\\
          \noalign{\smallskip}
\hline
1998   &      1.4   &      9.6    &    15.0   &     13.8   &      9.5    &    16.1  &      18.5    &    16.1\\
1999   &      2.2   &     11.7    &    22.6   &     13.6   &     10.0    &    17.0  &      13.6    &     9.3\\
2000   &      1.8   &      9.8    &    21.9   &     16.0   &      7.9    &    13.2  &      17.6    &    11.8\\
2001   &      1.2   &     11.6    &    23.1   &     14.8   &     10.1    &    15.2  &      15.7    &     8.4\\
2002   &      3.2   &      9.0    &    20.2   &     14.5   &      7.0    &    15.4  &      19.3    &    11.4\\
2003   &      2.8   &     10.1    &    24.4   &     14.1   &      9.8    &    14.7  &      16.6    &     7.5\\
2004   &      3.9   &      8.3    &    17.1   &     13.8   &      8.0    &    16.2  &      20.6    &    12.1\\
\hline
\hline
         \end{tabular}
         \end{center}
\end{table}
\begin{table}[t!]
     \begin{center}\scriptsize
       \caption[]{1985-2004 nighttime wind direction percentages for CAMC.}
    \label{camc-night-wdir}
         \begin{tabular}{r | c | c | c | c | c | c | c | c}
	 \hline\hline
          \noalign{\smallskip}
           & N & NE & E & SE & S & SW & W & NW\\
          \noalign{\smallskip}
\hline
1985 & 14.2 & 11.9 &  7.6 &  8.7 & 23.2 & 16.1 &  5.6 & 12.6\\
1986 & 22.5 & 18.4 & 12.9 & 12.9 & 12.4 &  7.5 &  5.5 &  8.0\\
1987 &  7.9 & 12.8 & 13.5 & 11.3 & 14.8 & 18.4 &  8.5 & 12.9\\
1988 &  7.6 &  9.6 & 16.6 & 14.3 & 20.1 & 12.3 &  6.6 & 12.9\\
1989 &  8.6 & 10.2 & 13.0 & 12.7 & 16.7 & 17.4 &  7.8 & 13.6\\
1990 &  7.6 & 11.5 & 16.5 & 13.5 & 15.0 & 17.8 &  7.6 & 10.5\\
1991 &  8.6 & 17.6 & 16.2 & 13.1 & 13.3 & 13.4 &  6.6 & 11.7\\
1992 &  9.3 & 14.4 & 18.0 & 16.6 & 13.7 &  9.5 &  7.8 & 10.7\\
1993 &  9.7 & 13.0 & 11.8 & 14.3 & 13.8 & 10.0 & 12.7 & 14.7\\
1994 & 26.6 & 13.6 &  9.1 & 12.5 & 15.5 &  8.7 &  5.8 &  8.2\\
1995 & 24.3 & 17.9 &  6.0 & 14.9 & 11.0 & 10.0 &  7.1 &  8.8\\
1996 &  5.4 &  1.7 &  6.8 & 47.1 & 17.3 &  7.1 &  7.7 &  6.7\\
1997 &  2.6 &  5.1 &  4.7 & 11.6 & 15.9 & 16.7 & 18.2 & 25.3\\
1998 &  2.0 &  4.8 &  4.2 & 11.3 & 16.1 & 20.4 & 22.6 & 18.7\\
1999 &  2.6 &  5.1 &  4.1 & 11.1 & 16.9 & 17.4 & 20.3 & 22.6\\
2000 &  7.8 &  6.8 &  5.3 & 11.3 & 15.0 & 16.9 & 18.5 & 18.5\\
2001 & 15.4 & 15.3 &  8.2 & 11.0 & 13.7 & 16.6 & 10.9 &  9.0\\
2002 & 15.5 & 14.2 &  9.1 & 16.3 & 16.7 & 12.2 &  8.9 &  7.2\\
2003 & 16.2 & 16.7 &  8.4 & 12.4 & 12.2 & 12.5 & 11.1 & 10.5\\
2004 & 71.8 &  3.6 &  2.6 &  5.6 &  7.9 &  4.1 &  1.7 &  2.6\\
\hline
\hline
         \end{tabular}
         \end{center}
\end{table}
\begin{table}[b]
     \begin{center}\scriptsize
       \caption[]{TNG, CAMC and NOT nighttime wind statistics (1998-2004).}
    \label{night-wsp}
         \begin{tabular}{l | r | r | r }
	 \hline\hline
          \noalign{\smallskip}
           & TNG & CAMC & NOT \\
          \noalign{\smallskip}
\hline
 $w_{sp}<3.3$ & 30.2\% & 83.6\% & 18.5\%\\
 $3.3\leq w_{sp}<12$ & 68.4\% & 16.4\% & 70.2\%\\
 $12\leq w_{sp}<15$ & 1.1\% &  0.0\% &  7.1\%\\
 $w_{sp}\geq 15$ & 0.3\% &  0.0\% & 4.2\%\\
\hline
\hline
         \end{tabular}
         \end{center}
\end{table}
Table \ref{night-wsp} shows the percentage of time per sector computed for four wind speed intervals. The last bin ($w_{sp}\geq 15$ m s$^{-1}$) is imposed by the safety observing conditions of TNG. Table \ref{night-wsp} shows that TNG and NOT have about 70\% of time in optimal wind speed conditions with respect to the 16.4\% of CAMC.\\
The evaluation of total time in which $w_{sp} > 15$ m s$^{-1}$ gives an estimation of the downtime due to high wind velocity. The lost time at TNG due to $w_{sp} > 15$ m s$^{-1}$ is only the 0.3\% of the total time. CAMC never shows $w_{sp} > 12$ m s$^{-1}$ and NOT is more affected by high wind speed (4.2\%).
\begin{table}[t]
     \begin{center}\scriptsize
       \caption[]{TNG, CAMC and NOT annual $RH$ percentages.}
    \label{rh-table}
         \begin{tabular}{r | c | c | c | c | c | c | c }
	 \hline\hline
          \noalign{\smallskip}
          Year & 1985 & 1986 & 1987 & 1988 & 1989 & 1990 & 1991\\
          \noalign{\smallskip}
\hline
TNG & $-$ & $-$ & $-$ & $-$ & $-$ & $-$ & $-$ \\
CAMC & 43.6 & 42.4 & 47.9 & 42.5 & 45.6 & 47.4 & 42.9\\
NOT & $-$ & $-$ & $-$ & $-$ & $-$ & $-$ & $-$\\
\hline
\hline
\noalign{\smallskip}
Year & 1992 & 1993 & 1994 & 1995 & 1996 & 1997 & 1998\\
\noalign{\smallskip}
\hline
TNG & $-$ & $-$ & $-$ & $-$ & $-$ & $-$ & 38.1\\
CAMC & 35.9 & 34.1 & 28.0 & 35.2 & 33.7 & 24.8 & 14.1\\
NOT & $-$ & $-$ & $-$ & $-$ & $-$ & $-$ & 39.4\\
\hline\hline
\noalign{\smallskip}
Year & 1999 & 2000 & 2001 & 2002 & 2003 & 2004 & 2005\\
\noalign{\smallskip}
\hline
TNG & 39.8 & 37.0 & 36.8 & 37.0 & 33.5 & 52.7 & 49.3\\
CAMC & 28.7 & 33.6 & 35.4 & 31.4 & 32.1 & 38.6 & $-$\\
NOT & 43.7 & 37.1 & 38.6 & 44.2 & 40.9 & 50.3 & 46.3\\
\hline\hline
         \end{tabular}
         \end{center}
\end{table}
\begin{table}[t]
     \begin{center}\scriptsize
       \caption[]{TNG, CAMC and NOT annual wintertime $RH$ percentages.}
    \label{rh-winter-table}
         \begin{tabular}{r | c | c | c | c | c | c | c}
	 \hline\hline
          \noalign{\smallskip}
          Year & 1985 & 1986 & 1987 & 1988 & 1989 & 1990 & 1991\\
          \noalign{\smallskip}
\hline
TNG & $-$ & $-$ & $-$ & $-$ & $-$ & $-$ & $-$\\
CAMC & 40.3 & 43.8 & 48.2 & 52.2 & 49.2 & 56.5 & 57.4\\
NOT & $-$ & $-$ & $-$ & $-$ & $-$ & $-$ & $-$\\
\hline
\hline
\noalign{\smallskip}
Year & 1992 & 1993 & 1994 & 1995 & 1996 & 1997 & 1998\\
\noalign{\smallskip}
\hline
TNG & $-$ & $-$ & $-$ & $-$ & $-$ & $-$ & 57.6\\
CAMC & 52.7 & 47.8 & 40.4 & 40.6 & 53.1 & 44.7 & 30.9\\
NOT & $-$ & $-$ & $-$ & $-$ & $-$ & $-$ & 61.1\\
\hline
\hline
\noalign{\smallskip}
Year & 1999 & 2000 & 2001 & 2002 & 2003 & 2004 & 2005\\
\noalign{\smallskip}
\hline
TNG & 43.2 & 54.2 & 37.6 & 58.4 & 39.0 & 44.7 & 75.0\\
CAMC & 12.5 & 53.0 & 33.7 & 56.4 & 36.3 & 40.8 & $-$\\
NOT & 54.7 & 58.1 & 36.8 & 69.2 & 54.0 & 51.0 & 75.7\\
\hline\hline
         \end{tabular}
         \end{center}
\end{table}
\begin{table}[t]
     \begin{center}\scriptsize
       \caption[]{TNG, CAMC and NOT annual summertime $RH$ percentages.}
    \label{rh-summer-table}
         \begin{tabular}{r | c | c | c | c | c | c | c}
	 \hline\hline
          \noalign{\smallskip}
          Year & 1985 & 1986 & 1987 & 1988 & 1989 & 1990 & 1991\\
          \noalign{\smallskip}
\hline
TNG & $-$ & $-$ & $-$ & $-$ & $-$ & $-$ & $-$\\
CAMC & 43.1 & 42.0 & 43.6 & 30.8 & 35.9 & 42.5 & 29.2\\
NOT & $-$ & $-$ & $-$ & $-$ & $-$ & $-$ & $-$\\
\hline
\hline
\noalign{\smallskip}
Year & 1992 & 1993 & 1994 & 1995 & 1996 & 1997 & 1998\\
\noalign{\smallskip}
\hline
TNG & $-$ & $-$ & $-$ & $-$ & $-$ & $-$ & 26.7\\
CAMC & 26.1 & 21.0 & 16.7 & 26.5 & 15.0 & 15.6 & 6.5\\
NOT & $-$ & $-$ & $-$ & $-$ & $-$ & $-$ & 25.4\\
\hline
\hline
\noalign{\smallskip}
Year & 1999 & 2000 & 2001 & 2002 & 2003 & 2004 & 2005\\
\noalign{\smallskip}
\hline
TNG & 26.2 & 28.1 & 23.6 & 30.2 & 28.0 & 43.3 & 26.6\\
CAMC & 19.7 & 24.3 & 21.9 & 22.0 & 24.5 & 32.2 & $-$\\
NOT & 21.4 & 27.5 & 23.3 & 31.2 & 25.7 & 41.1 & 20.9\\
\hline\hline
         \end{tabular}
         \end{center}
\end{table}
\begin{figure}[t]
   \centering
   \includegraphics[width=8cm]{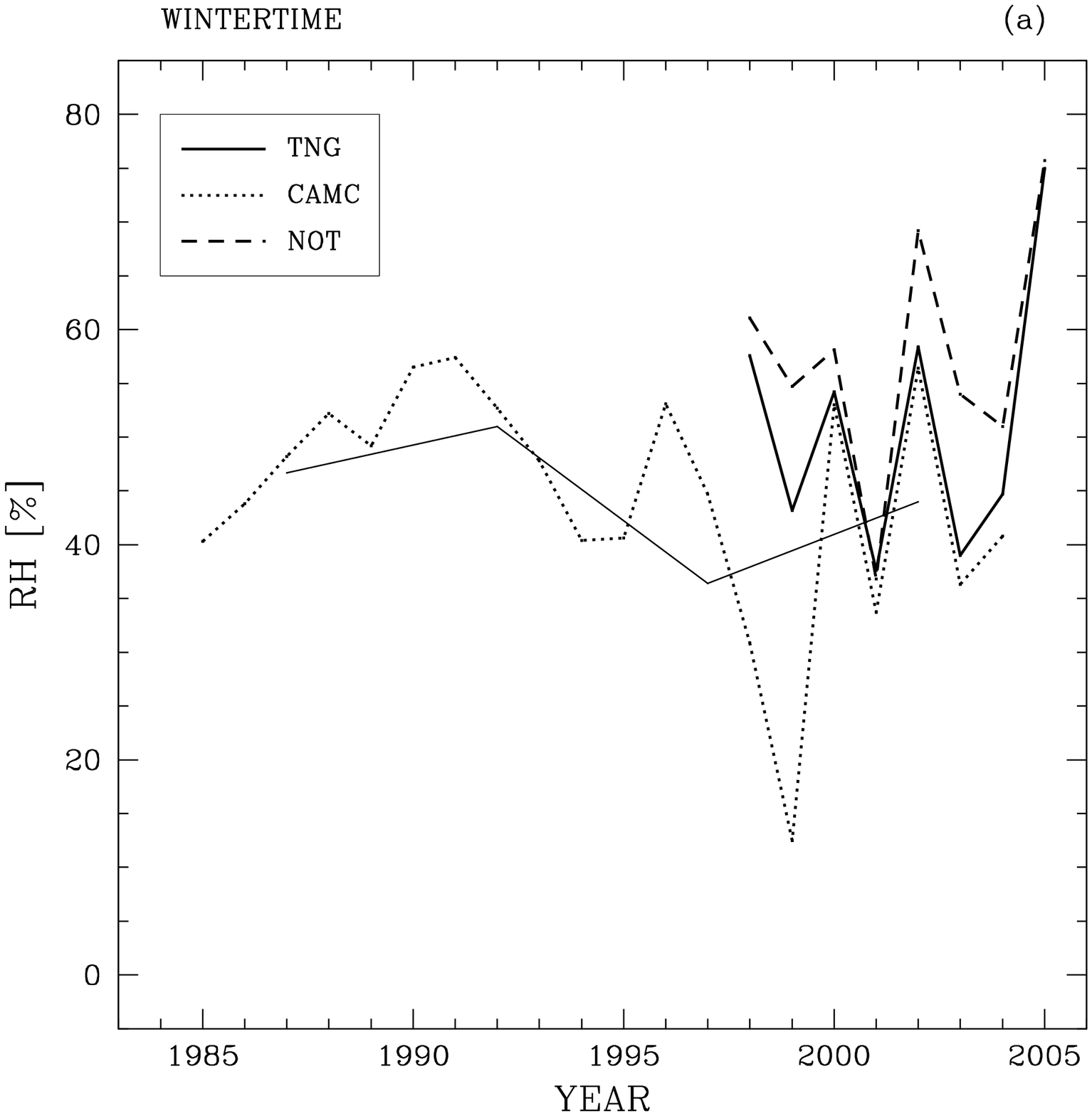}
   \includegraphics[width=8cm]{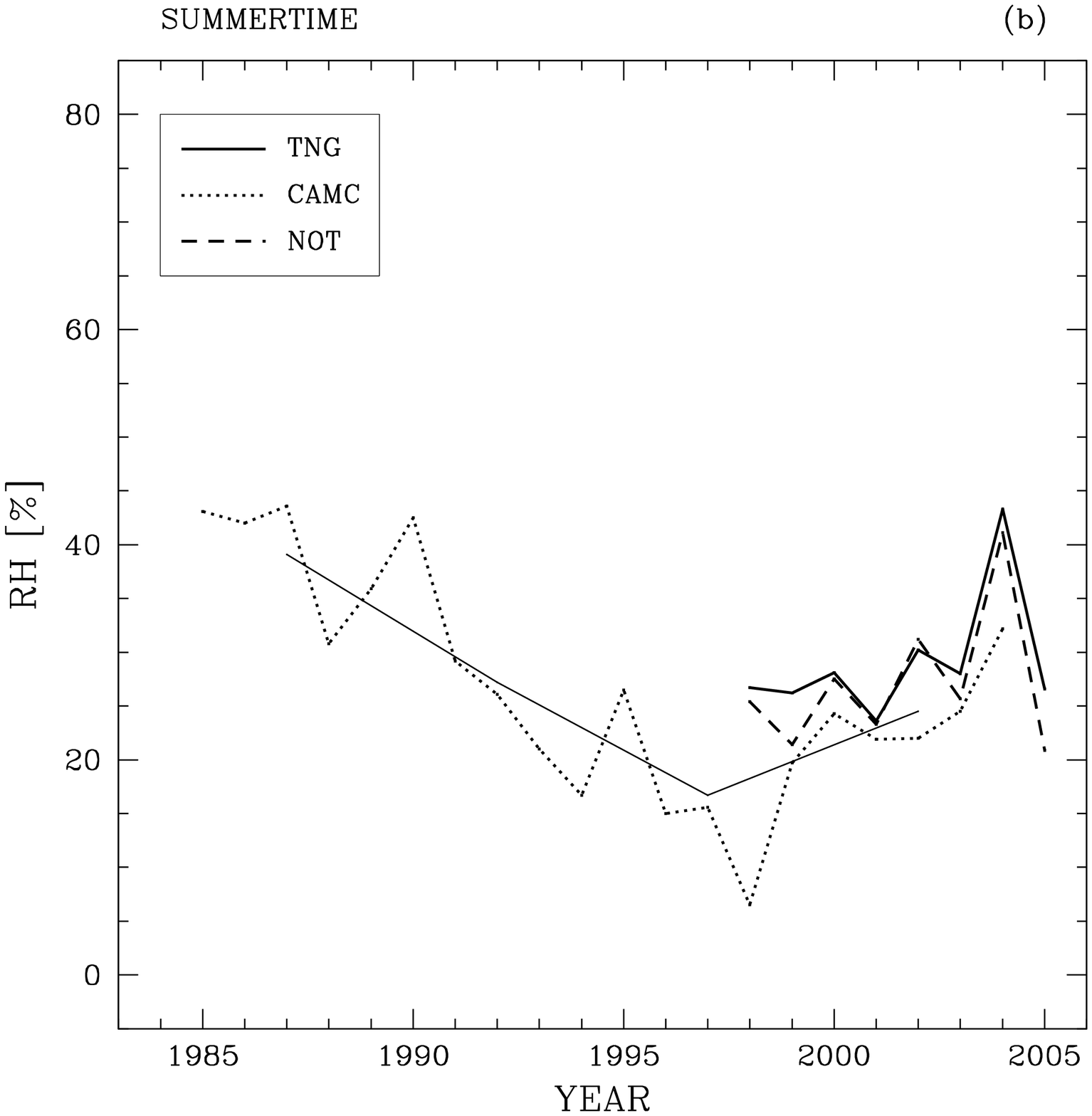}
   \caption{Annual $RH$ percentages at TNG (bold solid line), CAMC (dotted line) and NOT (dashed line) computed in wintertime (\textit{a}) and summertime (\textit{b}). The thin solid lines indicate the 5-years running mean of CAMC data series.}
              \label{rh-comparison}
    \end{figure}
\begin{figure}[t]
   \centering
   \includegraphics[width=8cm]{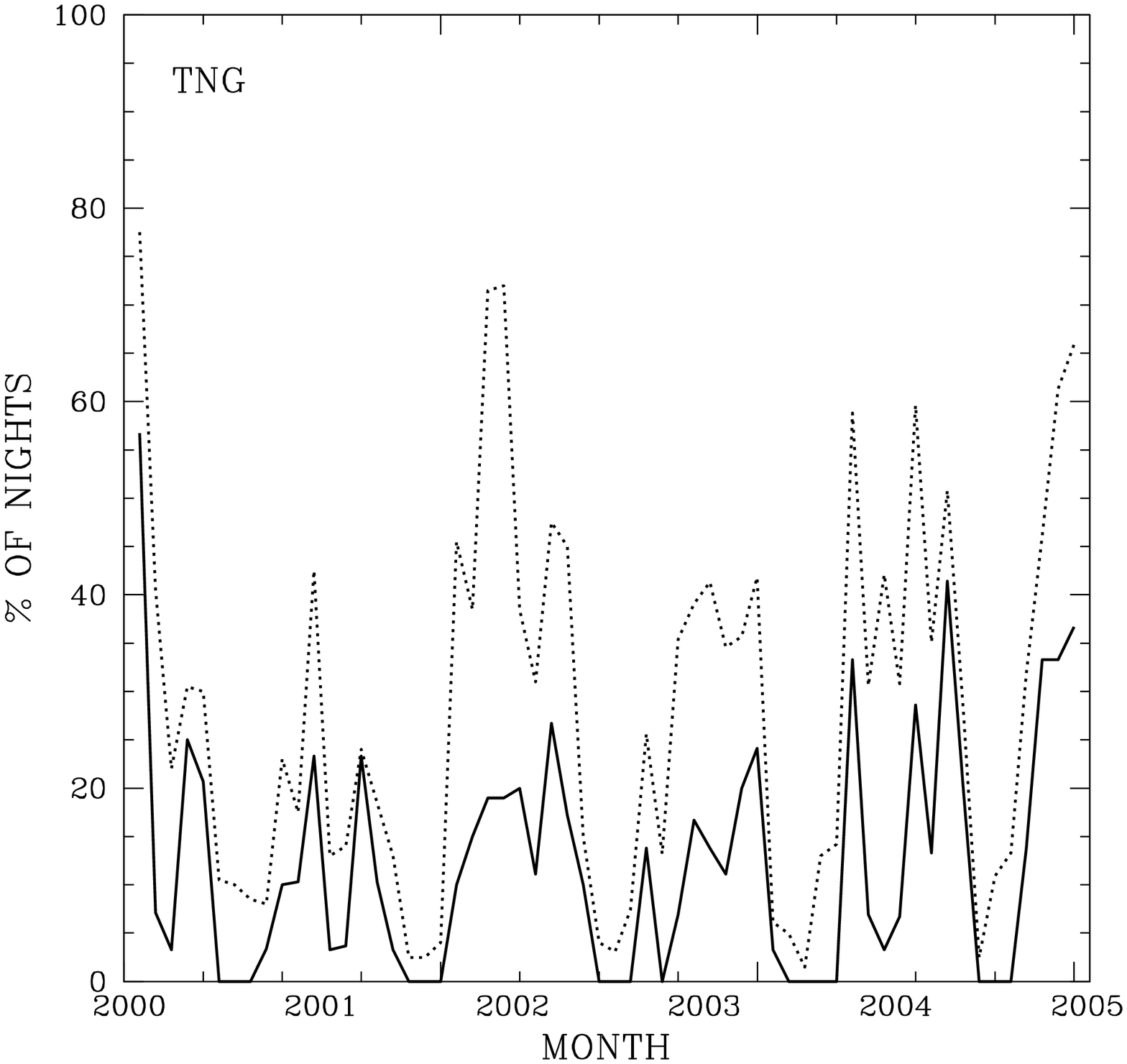}
   \includegraphics[width=8cm]{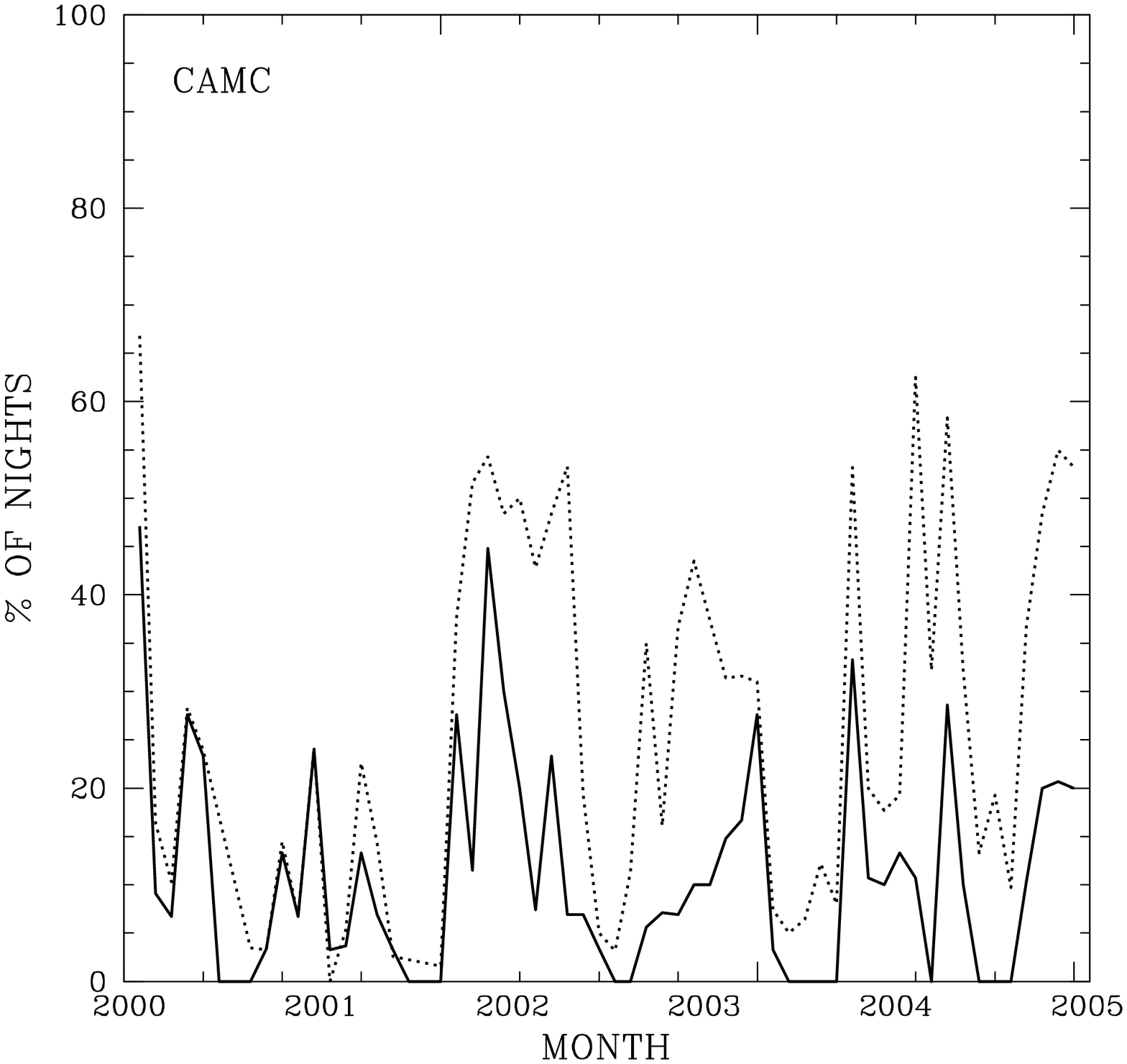}
   \caption{TNG (\textit{top}) and CAMC (\textit{bottom}) monthly percentages of nights with $RH>90\%$ (solid line) in comparison with monthly downtime (dotted line) in the period 2000-2005.}
              \label{monthly-downtime-90}
    \end{figure}
\begin{figure}[t]
   \centering
   \includegraphics[width=8cm]{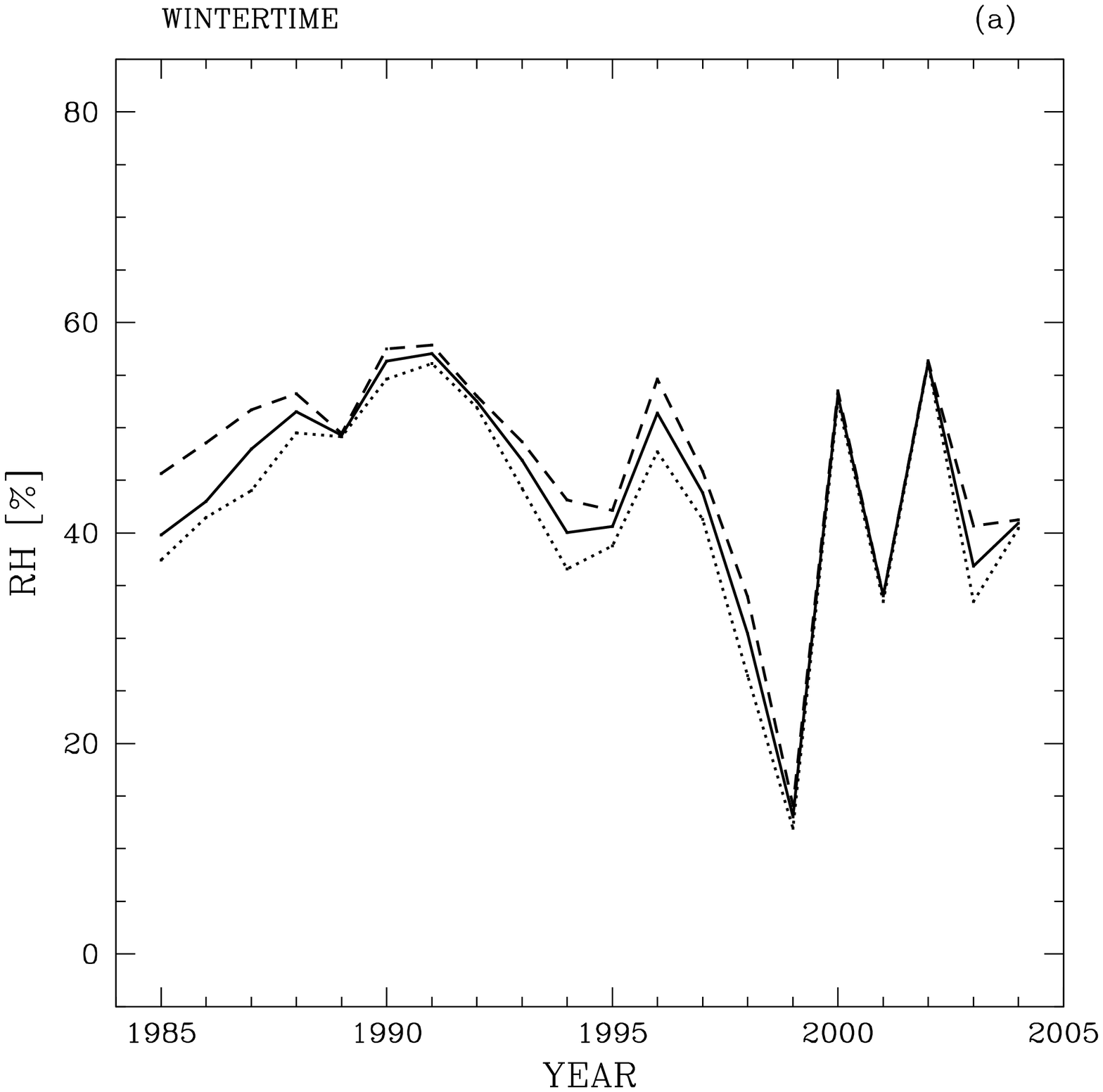}
   \includegraphics[width=8cm]{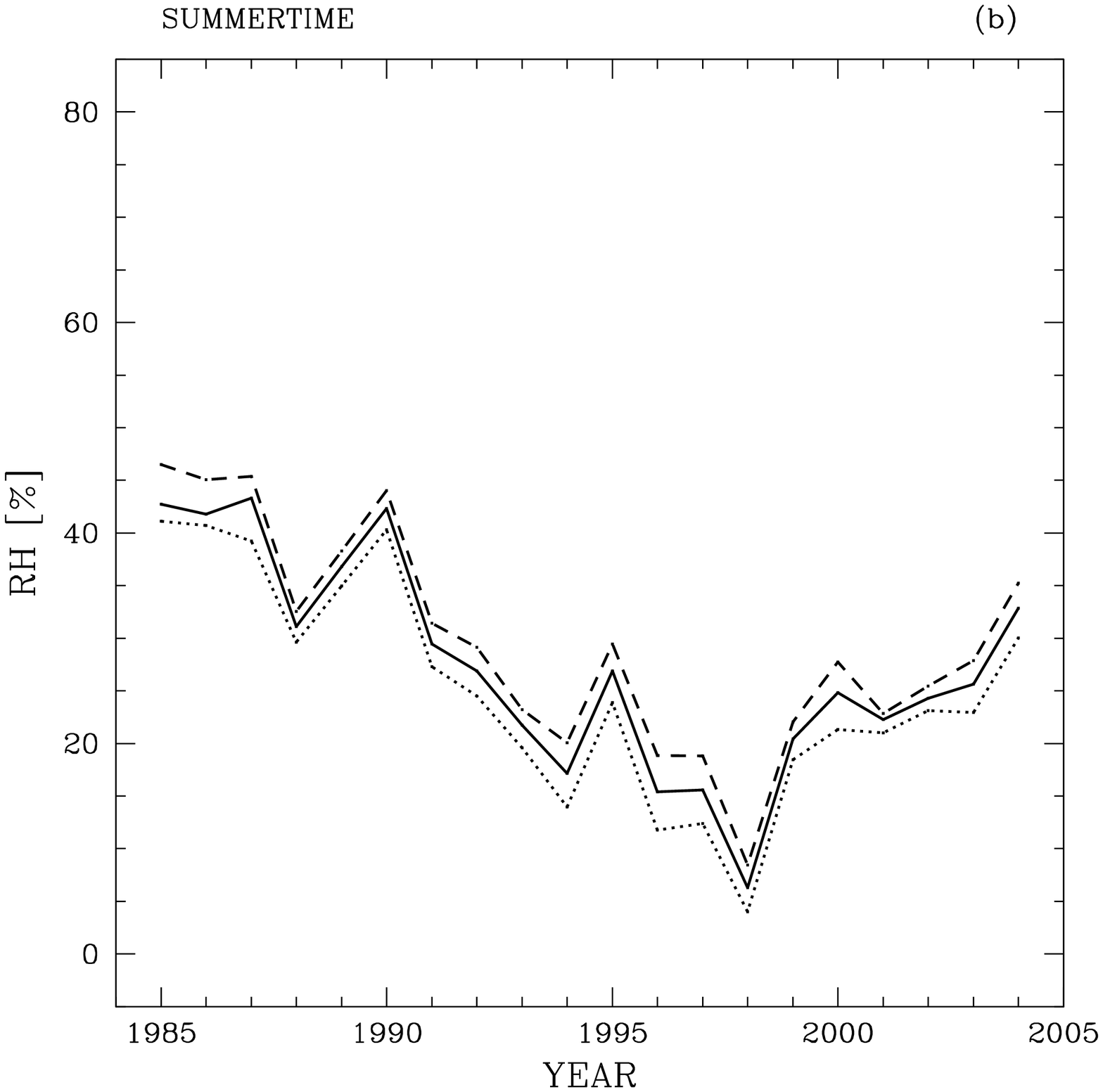}
   \caption{Comparison between CAMC annual daytime (dotted line), nighttime (dashed line) and entire day (solid line) $RH$ variations computed in wintertime (\textit{a}) and summertime (\textit{b}).}
              \label{camc-day-night-rh}
    \end{figure}
    
\section{Relative Humidity}\label{rh}
Table \ref{rh-table} reports the TNG, CAMC and NOT annual averages of $RH$ percentages, while Tables \ref{rh-winter-table} and \ref{rh-summer-table} report $RH$ percentages calculated in wintertime (October-March) and summertime (April-September) semesters for each year. Figure \ref{rh-comparison} shows the plot of the annual values as reported in Tables \ref{rh-winter-table} and \ref{rh-summer-table}. CAMC has the driest site mantaining a $RH<58\%$ in wintertime and a $RH<44\%$ in summertime, while both TNG and NOT have comparable trends and appear damper of $\sim 15\%$ on average in wintertime and $\sim 7\%$ on average in summertime. The 5-years running means of CAMC data series show a probable changing in slope after 1997. Strong anticorrelation between annual $RH$ and temperature trends has been found. The Spearman test gives a confidence level $> 99\%$.\\
High percentages of relative humidity make observing dangerous for the instrumentation. Optical surfaces become wet and can be damaged. In their statistics Murdin \cite{murdin} and Jabiri et al. \cite{jabiri} have used a threshold of 90\% of relative humidity. The percentage of nights in which $RH>90\%$ can be assumed as one of the main contributions to the total downtime of a telescope. In Figure \ref{monthly-downtime-90} the comparisons between monthly percentages of downtime and number of nights with $RH>90\%$ in the period 2000-2005 for TNG and CAMC are shown. It is intersting to note that the number of wet nights is a significant part of the total downtime, and the contribution of these nights vary with the season: it is higher in winter than in the summer.\\
In Table \ref{rh90} the annual percentages of nights with $RH>90\%$ for TNG, CAMC and NOT for the period 1998-2004 are shown. NOT shows the highest percentages in each year and TNG is only slightly lower, while CAMC has significant lower percentages.\\
\begin{table}[b]
     \begin{center}\scriptsize
       \caption[]{Annual percentages of nights with $RH>90\%$ in the period 1998-2004.}
    \label{rh90}
         \begin{tabular}{l | c | c | c | c | c | c | c}
	 \hline\hline
          \noalign{\smallskip}
           & 1998 & 1999 & 2000 & 2001 & 2002 & 2003 & 2004.\\
          \noalign{\smallskip}
\hline
 TNG & 13.1 & 17.0 &  13.3 &  8.9 & 10.2 & 9.7 & 18.9\\
 CAMC & 8.1 & 9.4 & 14.2 & 12.0 & 8.1 & 10.5 & 11.1\\
 NOT & 13.3 & 19.6 & 16.3 & 16.7 & 15.6 & 16.0 & 18.6\\
\hline
\hline
         \end{tabular}
         \end{center}
\end{table}
Our results have to be interpreted as lower limits of contributions because each telescope staff can assume its own safety standard procedures for what concerns the meteorological parameters.\\
In Figure \ref{camc-day-night-rh}-\textit{a} daytime and nighttime trends of wintertime $RH$ for CAMC are shown, while summertime trends are reported in Figure \ref{camc-day-night-rh}-\textit{b}. Figure \ref{camc-day-night-rh} shows that $RH$ have higher variations in wintertime with respect to summertime. In both cases nighttime $RH$ is higher with respect to daytime. $RH$ would simply change in consequence of the temperature changes due to the different heights of the Sun during the day. $RH$ is higher in the first hours of the morning (the coldest part of a day) than in the afternoon (the hottest part of the day). Irregular changes of temperature and relative humidity should be due to air masses movements and arrival of frontal systems.\\
\begin{figure}[t]
   \centering
   \includegraphics[width=8cm]{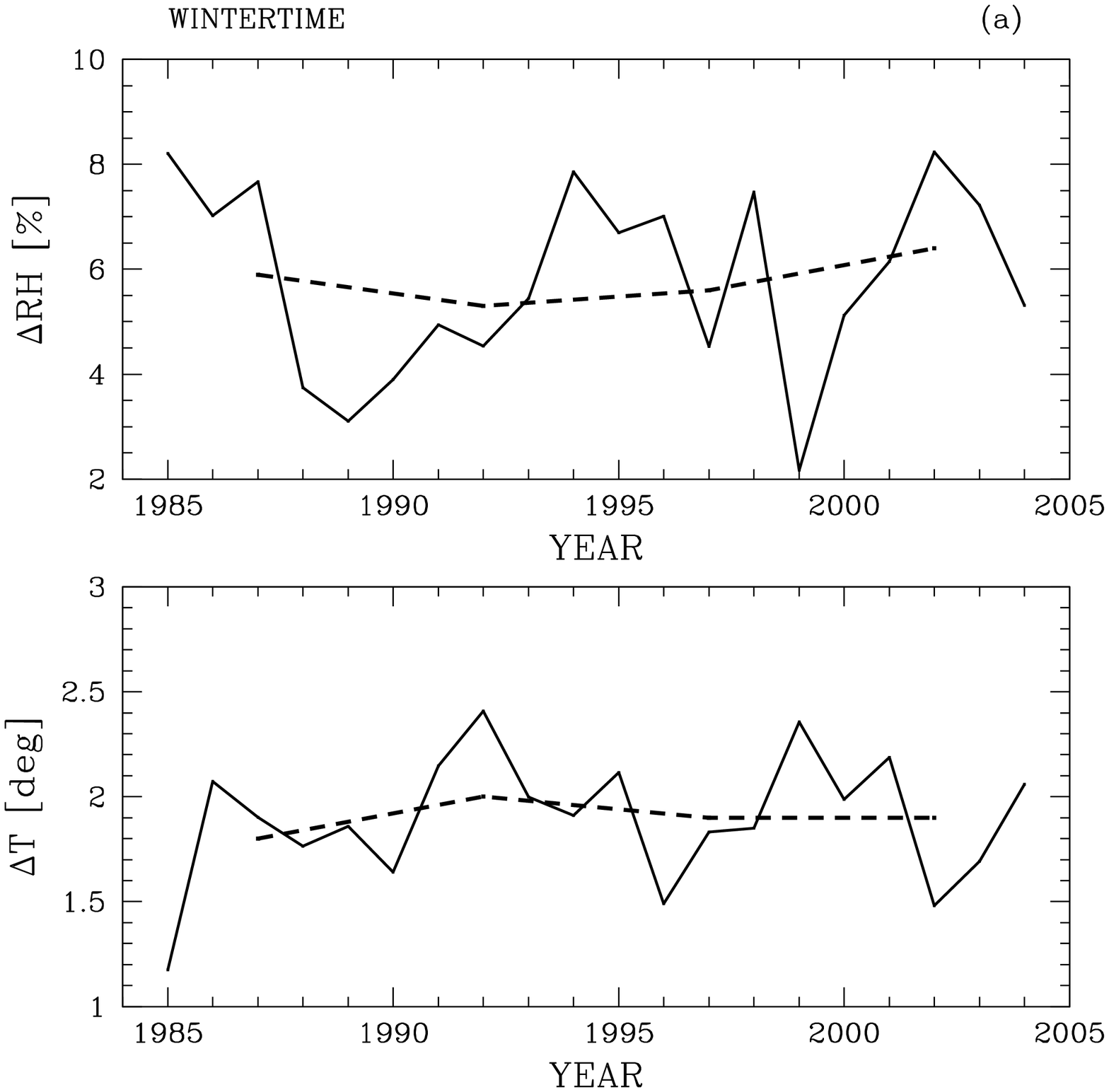}
   \includegraphics[width=8cm]{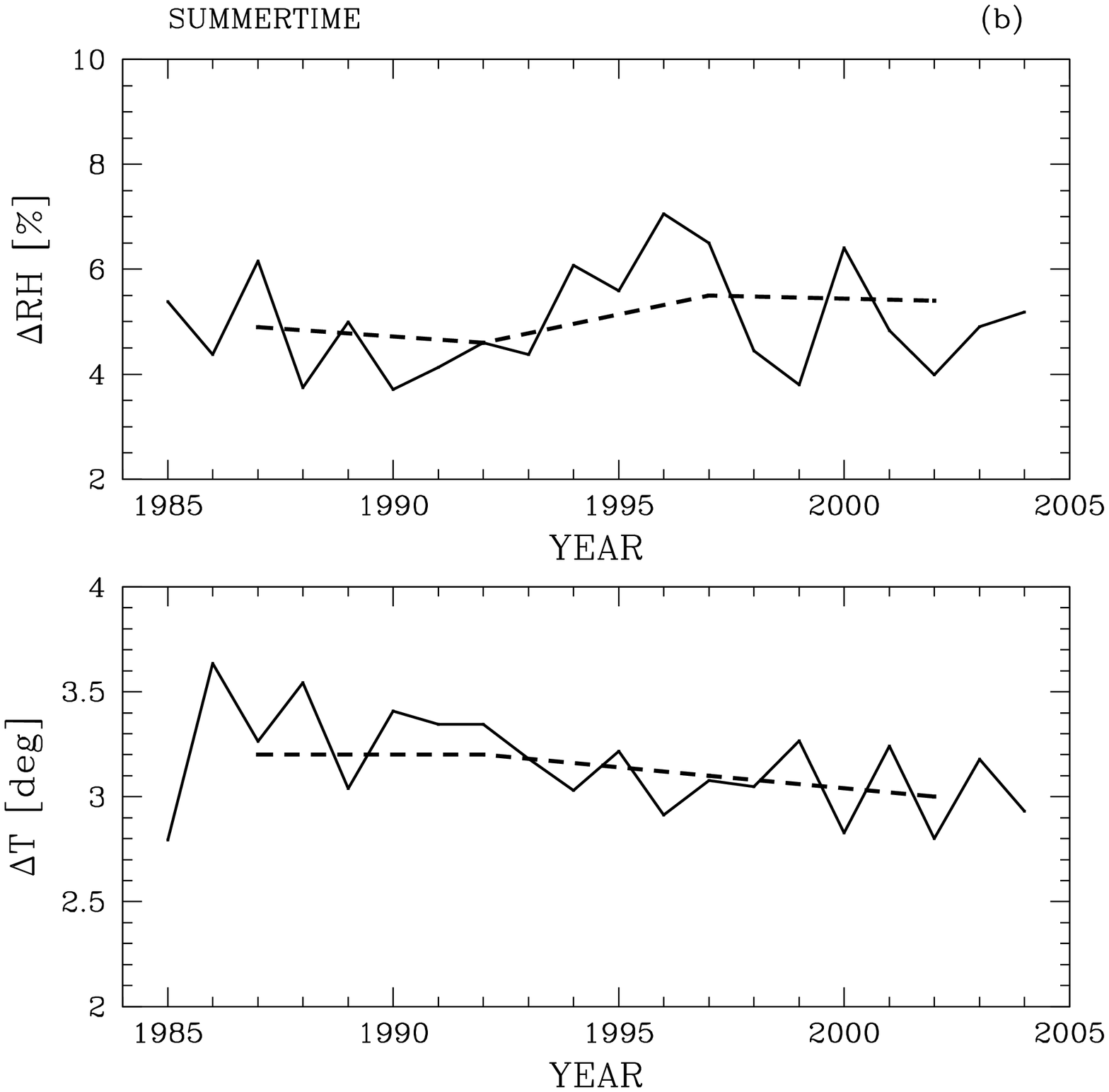}
   \caption{(\textit{a}) and (\textit{b}) \textit{Top}: CAMC trend of the differences between annual averages of daytime and nighttime $RH$. (\textit{a}) and (\textit{b}) \textit{Bottom}: CAMC trend of the differences between annual averages of daytime and nighttime $T$. The dashed lines indicate the 5-years running means. Plots (\textit{a}) are calculated in wintertime, plots (\textit{b}) in summertime.}
              \label{camc-rh-day-night}
    \end{figure}
The annual averages of the differences between daytime and nighttime $RH$ ($\Delta RH$) have been computed and the results for wintertime are reported in Figure \ref{camc-rh-day-night}-\textit{a} (\textit{top}) and for summertime in Figure \ref{camc-rh-day-night}-\textit{b} (\textit{top}). We note an oscillation of the values between 2\% and 8\% in wintertime and between 4\% and 7\% in summertime. Moreover, the 5-years running means represented by the dashed lines show an increasing trend in wintertime and a more steady trend in summertime.\\
In Figure \ref{camc-rh-day-night}-\textit{a} (\textit{bottom}) wintertime annual averages of the differences between daytime and nighttime $T$ ($\Delta T$) are shown, while summertime $\Delta T$ are reported in \ref{camc-rh-day-night}-\textit{b} (\textit{bottom}). $\Delta RH$ and $\Delta T$ are characterized by opposite trends and show a stronger anticorrelation in summertime (c.l. $\sim 96\%$) then in wintertime (c.l. $\sim 92\%$).\\

\section{Air Pressure}\label{pressure}
\subsection{Long period air pressure and weather conditions at ORM}\label{long-pressure}
\begin{figure}[t]
   \centering
   \includegraphics[width=8cm]{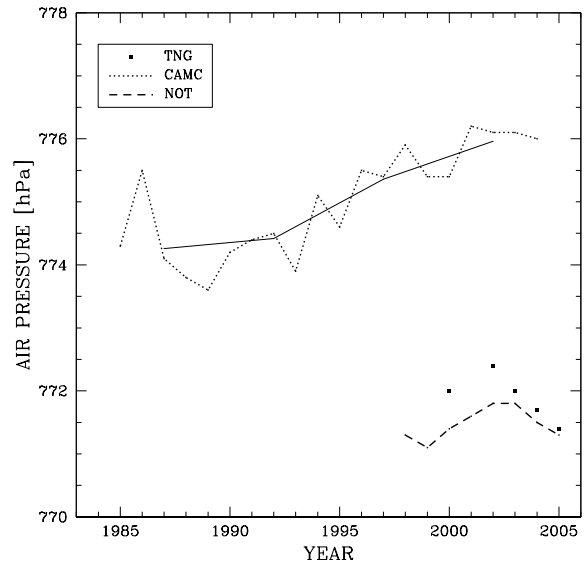}
   \caption{Annual $P$ averages at TNG (dots), CAMC (dotted line) and NOT (dashed line). The thin solid
   line indicates the 5-years running mean calculated on CAMC data series.}
              \label{p-comparison}
    \end{figure}

Table \ref{p-table} reports the TNG, CAMC and NOT annual averages of $P$. Since TNG database frequently suffers missing of pressure data, we have not obtained statistically significant annual averages for the years 1998, 1999 and 2001, then their annual mean values do not appear in Table \ref{p-table}. Figure \ref{p-comparison} shows the plot of the annual values reported in Table \ref{p-table}. As expected CAMC has the highest pressures (the range is between 773 and 776 hPa). It shows an increasing trend through the sampled 20 years, while NOT shows lower values in a range between 771 and 772 hPa. TNG annual averages have big differences with respect to NOT in the years 2000 and 2002, but show very similar values in the years 2003, 2004 and 2005. Table \ref{p-min-max} shows the maximum and minimum pressure values ever measured at the three telescopes. The difference between the absolute minimum and the absolute maximums is of about 38 hPa for TNG and CAMC, while is of about 26 hPa for NOT. All the absolute minimums have been measured in February 2004, while the absolute maximums have been registered in July 2001.\\
$P$ annual averages from CAMC show a strong correlation with temperature trends (c.l. $> 99\%$) and show an anticorrelation with the computed $RH$ annual averages (c.l. $> 99\%$).\\
The barometric correction due to the height of the three telescopes allows us to understand the differences in the mean values. We have thus investigated the exact altitude of the sensors. Caporali \& Barbieri \cite{caporali} have calculated the astronomical and geodetic coordinates of TNG using a class 1 electronic theodolite and a 12 channel GPS receiver NovAtel 3051. From such document we know that the optical axis of TNG is located at 2378 m above sea level. The difference in height between the optical axis and the sensor is about 22 m, so we obtain that the TNG sensor is located at an altitude $z_{TNG}\cong 2356$ m. From topographic maps we know that CAMC enclosure is at 2326 m above sea level, considering that the sensor is 1 m above the dome floor (cfr. \S \ref{data-analysis}) we have $z_{CAMC}\cong 2327$ m. Finally, the NOT dome is at 2382 m and the sensor at 2 m above the dome floor (cfr. \S \ref{data-analysis}), giving $z_{NOT}\cong 2384$ m.\\
\begin{table}[t]
     \begin{center}\scriptsize
       \caption[]{TNG, CAMC and NOT annual $P$ averages [hPa].}
    \label{p-table}
         \begin{tabular}{r | c | c | c | c | c | c | c}
	 \hline\hline
          \noalign{\smallskip}
          Year & 1985 & 1986 & 1987 & 1988 & 1989 & 1990 & 1991\\
          \noalign{\smallskip}
\hline
TNG & $-$ & $-$ & $-$ & $-$ & $-$ & $-$ & $-$\\
CAMC & 774.3 & 775.5 & 774.1 & 773.8 & 773.6 & 774.2 & 774.4\\
NOT & $-$ & $-$ & $-$ & $-$ & $-$ & $-$ & $-$\\
\hline
\hline
\noalign{\smallskip}
Year & 1992 & 1993 & 1994 & 1995 & 1996 & 1997 & 1998\\
\noalign{\smallskip}
\hline
TNG & $-$ & $-$ & $-$ & $-$ & $-$ & $-$ & $-$\\
CAMC & 774.5 & 773.9 & 775.1 & 774.6 & 775.5 & 775.4 & 775.9\\
NOT & $-$ & $-$ & $-$ & $-$ & $-$ & $-$ & 771.3\\
\hline
\hline
\noalign{\smallskip}
Year & 1999 & 2000 & 2001 & 2002 & 2003 & 2004 & 2005\\
\noalign{\smallskip}
\hline
TNG & $-$ & 772.0 & $-$ & 772.4 & 772.0 & 771.7 & 771.4\\
CAMC & 775.4 & 775.4 & 776.2 & 776.1 & 776.1 & 776.0 & $-$\\
NOT & 771.1 & 771.4 & 771.6 & 771.8 & 771.8 & 771.5 & 771.3\\
\hline\hline
         \end{tabular}
         \end{center}
\end{table}
\begin{table}[b]
     \begin{center}\scriptsize
       \caption[]{Maximum and minimums pressure values at TNG, CAMC and NOT.}
    \label{p-min-max}
         \begin{tabular}{l | c | c | c}
	 \hline\hline
          \noalign{\smallskip}
           & Min & Max & Difference\\
          \noalign{\smallskip}
\hline
 TNG & 743.1 & 781.9 & 38.8\\
 CAMC & 747.3 & 785.8 & 38.5\\
 NOT & 752.1 & 778.4 & 26.3\\
\hline
\hline
         \end{tabular}
         \end{center}
\end{table}
We can compute the theoretical pressure for each site height using the barometric correction that depends on site's scale height ($H$) in the barometric law, which is dependent on the mean temperature $\left\langle T_{layer}\right\rangle$ of the layer in which the correction is applied. TNG and CAMC annual temperatures in the period 1998-2004 give a mean temperature of $9.5 \pm 0.5$ deg that can be used as mean temperature at ORM. Graham (2005, unpublished) gives a mean temperature of about $20.3 \pm 0.4$ deg at Mazo Airport, located few meters above sea level at La Palma Island. The mean temperatures at ORM and at Mazo Airport allow us to calculate the mean temperature of the layer between the sea level at La Palma and the ORM using the weighted average of the two measures. We obtain $\left\langle T_{layer}\right\rangle = 16.1 \pm 0.3$ deg.\\
The closest $\left\langle T_{layer}\right\rangle$ value in the standard atmospheric model of Glenn Research Center at NASA web site\footnote{http://www.grc.nasa.gov} is 15 deg and gives a theoretical pressure value of 763.3 hPa for an altitude of 2327 m (altitude of the CAMC sensor), which corresponds to a theoretical scale height $H_{GRC} \cong 8220$ m. If we use standard tables in Allen \cite{allen}, still for $\left\langle T_{layer}\right\rangle = 15$ deg, we found a theoretical scale height $H_{Allen} \cong 8430$ m. $H_{GRC}$, $H_{Allen}$ are in good agreement and give a mean value of $H = 8325$ m.\\
Once $H$ is fixed, the barometric law gives us the standard theoretical pressures scaled at the altitudes of TNG, CAMC and NOT sensors: $P_{TNG} = 763.3$ hPa, $P_{CAMC} = 766.0$ hPa and $P_{NOT} = 760.8$ hPa. All these thoretical results are lower than the respective pressures of Table \ref{p-table}, confirming that ORM is dominated by high pressure. As is to be expected from anticyclonic conditions, this means prevailing stable good weather.\\
Once the barometric correction is applied to the empiric results in Table \ref{p-table}, the mean pressure differences between the three telescopes is of about 1 hPa. This value can be assumed as the upper limit of the error deriving from the barometric correction. 
\subsection{Short period air pressure and temperature correlations}\label{short-pressure}
\begin{figure}[t]
   \centering
   \includegraphics[width=8cm]{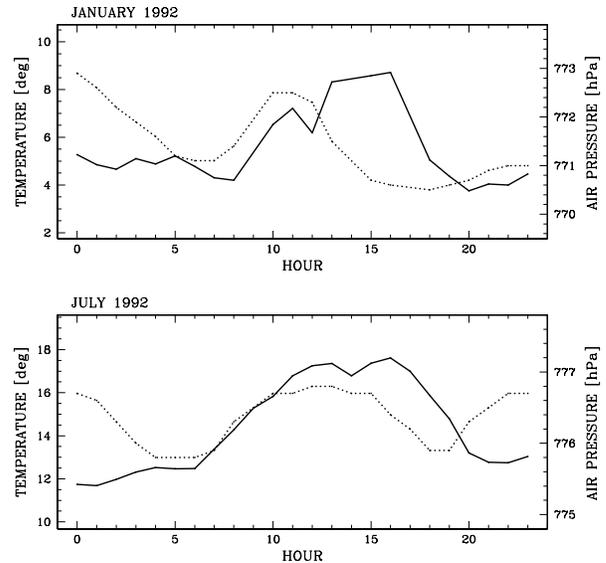}
   \caption{Hour-by-hour $P$ (dotted line) and $T$ (solid line) trends at CAMC for year 1992.}
              \label{hour-tp}
    \end{figure}

\begin{figure}[t]
   \centering
   \includegraphics[width=8cm]{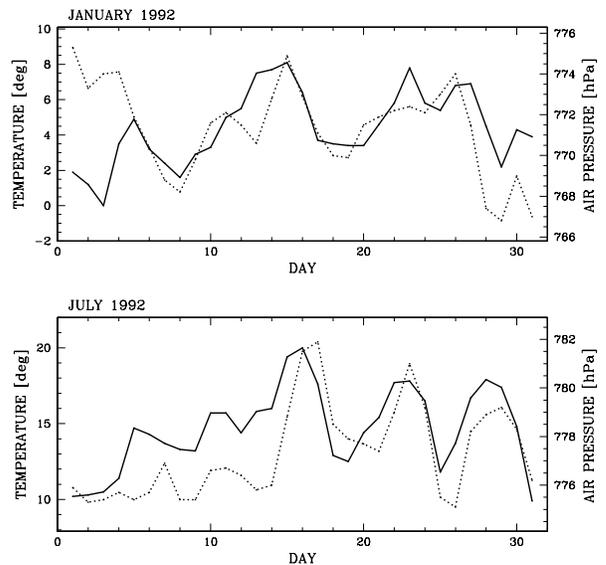}
   \caption{Day-by-day $P$ (dotted line) and $T$ (solid line) trends at CAMC for year 1992.}
              \label{day-tp}
    \end{figure}

We have also analyzed short time scales relationships (hour-by-hour and day-by-day) between air pressure and temperature variations in wintertime (January) and summertime (July) for several years. Examples for year 1992 at CAMC are reported in Figure \ref{hour-tp} and Figure \ref{day-tp}.\\
For what concerns hour-by-hour relationship, we have taken into account single day trends of $P$ and $T$ hourly averages among 24 hours. Figures \ref{hour-tp} clearly shows that pressure anticipates the temperature behaviour tipically by 2-3 hours both in wintertime and summertime. This result can be read in another way: the arrival of warmer air masses induces the collapse of pressure in correspondence of the increase of temperature.\\
The relationship between $P$ and $T$ suggests that it is possible to foresee with 2-3 hours in advance the changes of the temperature on the basis of the changes in pressure. In this way it is possible to optimize the thermalization of the telescope and of the instruments reducing the instrumental seeing. The correlation between pressure and respective temperature measured 2 hours later has confidence level $> 98\%$. Tipically this correlation decreases if temperatures are measured 1 hour (c.l. $< 84\%$) or 3 hours (c.l. $\sim 95\%$) after the relative correlated pressure.\\
Day-by-day correlations between $P$ and $T$ have been also analyzed taking into account the daily averages for an entire month. In this case the link between pressure and temperature becomes less evident (Figure \ref{day-tp}) and disappears in longer time scales. This means that the capacity of prediction based on the hour-by-hour analysis vanishes in time scales higher than few hours. A more statistical analysis should be interesting in particular to correlate variations of pressure with the seeing to the aim to predict the optimal seeing conditions.\\
It is also interesting to note that the range between the measured maximum and minimum pressure ($\Delta P$) is higher in wintertime than in summertime for both hour-by-hour and day-by-day time scales.

\section{Conclusions}
In this paper we present for the first time an analysis of long term meteorological data directly obtained from local meteorological towers of TNG, CAMC and NOT, at a height of about 2300 m above sea level, well above the inversion layer.\\
Wind vector $\vec{V}$, splitted into scalar wind speed ($w_{sp}$) and vectorial wind direction ($\vec{w}_{dir}$), shows different behaviours for each telescope.\\
We also investigated the influence of wind speed on astronomical seeing at TNG, and we found that the seeing deteriorates when $w_{sp} < 3.3$ m s$^{-1}$. Sarazin \cite{sarazin92} shows in La Silla a limiting value of $w_{sp}<2$ m s$^{-1}$ and a $w_{MAX} = 12$ m s$^{-1}$. We can conclude that the percentage of time in which $w_{sp} \in$ [3.3, 12[ is the signature of wind speed dependent good seeing conditions. TNG and NOT have 70\% of time in this interval, while CAMC has only 16.4\%. The evaluation of total time in which $w_{sp} > 15$ m s$^{-1}$ gives an estimation of the downtime due to high wind speed. High wind speed at TNG results to be the 0.3\%  of the total time. CAMC never shows $w_{sp} > 12$ m s$^{-1}$ and NOT is more affected by high wind speed (4.2\%).\\
Relative humidity ($RH$) annual averages for the three telescopes have been calculated for entire year, wintertime and summertime. CAMC has the driest site, while both TNG and NOT have comparable trends and appear damper. The 5-years running means calculated on CAMC data series show a probable changing in slope after 1997. Strong anticorrelation between $RH$ and temperature trends has been found (c.l. $> 99\%$). The percentage of nights in which $RH>90\%$ can be assumed as one of the main contributions to the total downtime of a telescope. NOT has the highest percentages and TNG is only slightly lower, while CAMC has significant lower percentages. Wintertime and summertime annual averages of the differences between daytime and nighttime $RH$ ($\Delta RH$) oscillate with increasing trends anticorrelated with wintertime and summertime annual averages of differences between daytime and nighttime temperature ($\Delta T$). $\Delta RH$ and $\Delta T$ are characterized by opposite trends and show a stronger anticorrelation in summertime (c.l. $\sim 96\%$) then in wintertime (c.l. $\sim 92\%$).\\
Air pressure is correlated with temperature (c.l. $> 99\%$) and anticorrelated with annual relative humidity (c.l. $> 99\%$). The analysis of annual averages of air pressure has confirmed that ORM is dominated by high pressure, that means prevailing stable good weather. We have estimated the scale height of La Palma with theoretical models obtaining $H = 8325$ m. Short time scales relationships (hour-by-hour and day-by-day) between air pressure and temperature variations in wintertime and summertime have been analyzed too. Hour-by-hour analysis shows that pressure anticipates the temperature changes tipically by 2-3 hours both in wintertime and summertime and this suggests a possibility to foresee with 2-3 hours in advance the changes of the temperature on the basis of the changes in pressure (c.l. $> 98\%$). In this way it is possible to optimize the thermalization of the telescope and of the instruments reducing the instrumental seeing. This capacity vanishes in time scales higher than few hours because the link between pressure and temperature becomes less evident in the day-by-day analysis and disappear in longer time scales.

\begin{acknowledgements}
The authors acknowledge the CAMC and NOT staff for the availability of the meteorological data on-line and the anonymous reviewer for helpful comments. G. Lombardi thanks also Ernesto Oliva of TNG for the useful informations and data, Jose L.
Mui\~nos Haro of CAMC for his kindness, support with the data and the informations, Ricardo Javier C\'ardenas Medinas and Peter Meldgaard Sorensen of NOT for their help and the patience, A. Bragaglia for the useful images from TNG.
\end{acknowledgements}

\end{document}